\documentclass[fleqn,10pt]{wlscirep}

\usepackage[T1]{fontenc}
\usepackage{float}

\def\beq{\begin{equation}}
\def\eeq{\end{equation}}
\def\bea{\begin{eqnarray}}
\def\eea{\end{eqnarray}}

\newcommand{\chemN}{\texttt{N}\,}
\newcommand{\chemO}{\texttt{O}\,}
\newcommand{\chemNO}{\texttt{NO}\,}
\newcommand{\chemOtwo}{\texttt{O}$_2$\,}

\newcommand{\chemOzone}{\texttt{O}$_3$\,}
\newcommand{\chemNtwo}{\texttt{N}$_2$\,}
\newcommand{\chemNOtwo}{\texttt{NO}$_2$\,}

\newcommand{\chemNOy}{\texttt{NO}$_{y}$\,}

\newcommand{\IGNORE}[1]{}

\title{The Effects of Gamma Ray Bursts on Extinction and Survivability in the Galaxy 
}

\author[1,*,+]{Matan Sade}
\author[2]{Aviv Tsarfati}
\author[1,**,+]{Ofek Birnholtz}

\affil[1]{Department of Physics, Bar-Ilan University, Ramat Gan 5290002, Israel}
\affil[2]{Faculty of Biotechnology and Food Engineering, Technion –- Israel Institute of Technology, Haifa 3200003, Israel}
\affil[*]{sadeh.matan@live.biu.ac.il}
\affil[**]{ofek.birnholtz@biu.ac.il}
\affil[+]{these authors contributed equally to this work}

\begin{abstract}
High-energy astrophysical events, particularly Gamma Ray Bursts (GRBs), have been proposed as significant contributors to mass extinction events on Earth-like planets in most of the galaxy, internal to our radius in it.
This paper examines the extent to which GRBs may reset the evolutionary progress of complex life through repeated extinction-level disruptions.
While resilient extremophiles may survive even the most intense GRBs,
    more complex surface-dwelling organisms are vulnerable to indirect atmospheric effects,
    primarily UV exposure following ozone depletion.
By identifying evolutionary milestones and estimating how frequently GRBs would need to occur to prevent recovery between such milestones,
    this work proposes that GRBs could act as evolutionary filters,
    limiting the emergence of advanced life, but only much closer to the galactic center.
We consider the implications for searches of various biosignatures versus technosignatures.
\end{abstract}

\begin{document}
\raggedbottom
\maketitle
\setlength{\parindent}{0pt}


\begin{figure}[t]
\centering
\includegraphics[width=0.8\textwidth]{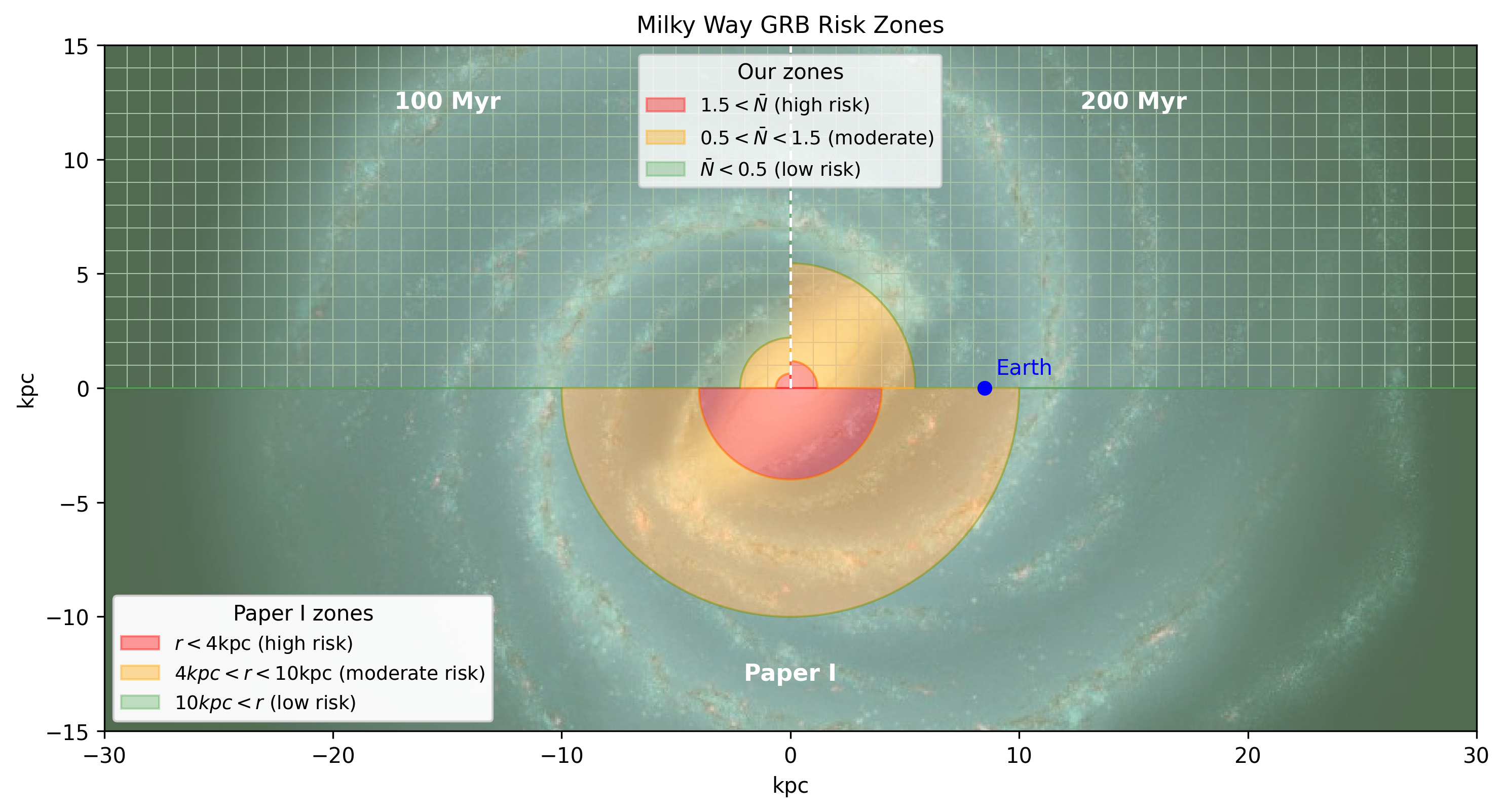}
    \caption{\textbf{Comparison of GRB risk-level zones in our galaxy, by radius.}
    The bottom half reproduces the estimates from Paper I.
    The top half shows our re-evaluated risk zones,
        derived from the shorter exposure windows.
    We present both the conservative (200 Myr, right)
    and aggressive (100 Myr, left) ancestral exposure windows,
    defined by the expected number of lethal GRBs $\bar{N}$ per window at each galactic radius.
    The Earth's galactic location is marked relative to a model background by NASA/JPL-Caltech/R. Hurt (SSC/Caltech) (ssc2008-10a).
    }
\label{fig:zones}
\end{figure}

\section*{Introduction}
\label{sec:1}
A debate has emerged in recent years \cite{Piran,Loeb} over the possible role of Gamma Ray Bursts (GRBs) \cite{Paczynski:1991aq,Piran:1999kx,Mukherjee,Barnacka,Savaglio} as an astrophysical cause of mass extinctions, and furthermore, as an existential threat to life on Earth-like planets.
Paper I \cite{Piran} specifically calculated that the GRB rate in the interior of our galaxy,
    even as far out radially as the Earth's location and beyond
    (see bottom of Fig. \ref{fig:zones}),
    is high enough to eradicate life,
    and thus serve as a potential explanation of the Fermi Paradox \cite{Sullivan,Drake,Sagan,Jones} of extraterrestrial life not being found. 
Conversely, Paper II \cite{Loeb} argued that GRBs pose no existential threat to life \textit{as a whole},
    since some exceptionally resilient lifeforms --
    namely extremophiles, such as \textit{Milnesium tardigradum} --
    would survive such GRBs,
    unless the GRB occurred so close that its direct power output were sufficient to boil the oceans —-
    well beyond expected GRB fluences at a relevant rate.
Both papers agree, to begin with,
    that for reasonably close GRBs,
    the most lethal effects of are not the direct radiation burst,
    but rather the atmospheric and ecological chain reactions that follow \cite{Thomas, Rampelotto}.
In this work, we seek to expand the understand of the damage mechanisms,
    and to assess their effects in the context of a planet like the Earth,
    analyzed as an evolving geological, ecological and biological system,
    and so to better refine the threat and its effects on advanced life.
We also broaden the focus from extremophiles to more general
    and more diverse
    complex lifeforms,
    looking into both damage and recovery.
We take the Earth as a case-study,
    and note throughout which assumptions or conditions might be generalizable.

In what follows, we first review the damage a GRB can be expected to cause to life on a planet like the Earth's present-day state,
    and map the GRB rate and galactic danger zone according to Paper I.
We then expand Paper II's point regarding likely survivors from a single extremophile species to general complex species,
    and relate the exposure of such species to Earth's bio-systematic timeline.
From these considerations we significantly diminish the danger zones,
    and conclude that GRBs do not provide a significant likely contribution to resolving the Fermi paradox.

\begin{figure}[t]
 \begin{center}
\includegraphics[width=0.8\textwidth]{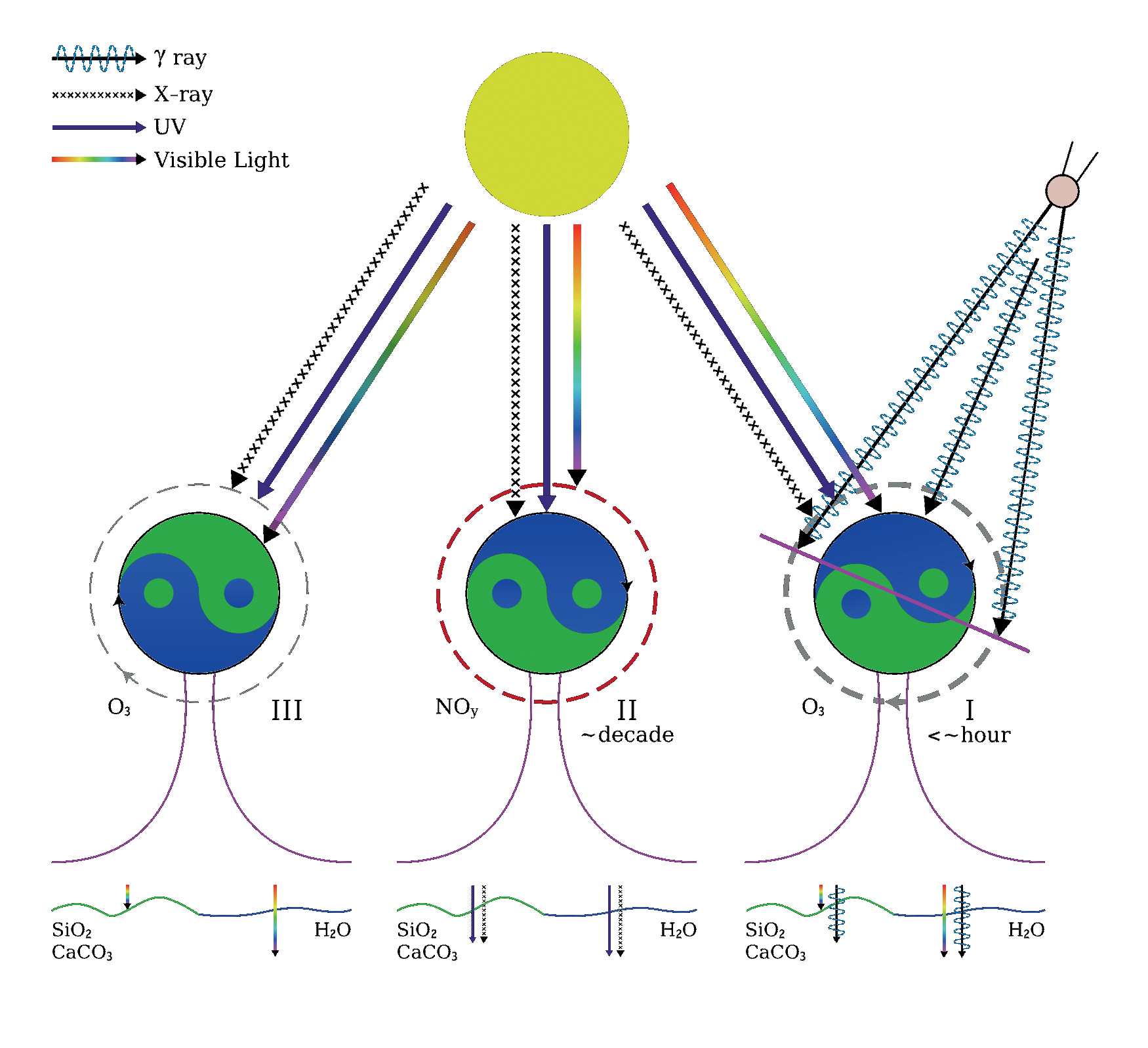}
\caption{The phenomenon, represented by three stages. Seen in the figure: a G-type main-sequence star (yellow), an Earth-like planet (blue-green yinyang), a GRB (beige), an ozone layer (gray), \chemNOy layer \cite{Perros} (red). The relevant electromagnetic radiation is shown in the legend found in the top-left of the figure. The penetration of electromagnetic radiation is shown qualitatively by the meeting of the radiation arrow with the layer (radiation is mostly returned/scattered) or the continuation of the radiation through the layer (radiation penetrates the layer).
}
\label{fig:stages}
\end{center}
\end{figure}

\section*{The Danger of GRBs}

The biological impact of GRBs can be modeled in three stages, adopted by Paper I\cite{Piran} from elsewhere \cite{Thomas} (see Fig. \ref{fig:stages}):

\textbf{Stage I} involves the direct emission of intense gamma radiation over a short time period.
GRBs can strike a planet from various angles \cite{Tegmark},
    and due to their brief duration relative to planetary rotation \cite{Stephenson}, typically only one hemisphere is directly exposed to the radiation.
The immediate biological damage in this stage is largely restricted to the irradiated hemisphere and is often considered insufficient to cause global extinction on its own.
However, the incident $\gamma$ rays interacting with the upper atmosphere also dissociate diatomic nitrogen,
\chemNtwo $\to$ 2\chemN,
and subsequent reactions with atmospheric oxygen form nitric oxide:
\chemN + \chemOtwo $\to$ \chemNO + \chemO.

\textbf{Stage II}, is the most significant in terms of long-term biological consequences,       as at this stage the Nitric oxide and related compounds (\chemNOy\!) act catalytically to destroy ozone, 
    \chemNO + \chemOzone $\to$ \chemNOtwo + \chemOtwo.
While this is a simplified chain,
    the reactions above are representative of the dominant mechanisms;
    a full treatment can be found in \cite{Thomas}.
This is also a global effect
    (though not necessarily uniform),
    due to atmospheric mixing.
The depletion of the ozone layer (see Table \ref{table:grbs}, adapted from Paper I) permits harmful UVB and UVC radiation to reach the surface,
    posing severe risks to photosynthetic organisms, shallow marine life, terrestrial ecosystems, and any species dependent on surface conditions.

\textbf{Stage III} begins once atmospheric chemistry and transport processes restore the ozone layer,
    typically (on Earth simulations) over the course of about a decade \cite{Thomas}.
Although ozone begins recovering shortly after the event,
    the biological consequences from sustained high-UV exposure can extend throughout this period \cite{Thomas}.
The end of Stage III marks a return of atmospheric conditions to Stage I,
    albeit with possible great changes in the surviving species composition.

\begin{table}[h!]
\centering
\begin{tabular}{|p{3.5cm} p{2.2cm} p{2.2cm} p{2.2cm} p{2cm}|}
 \hline\hline
 Fluence (LGRB) 
    & $P(t < 5\, \text{Gyr})$
    & $P(t < 1\, \text{Gyr})$
    & $P(t < 0.5\, \text{Gyr})$
    & \chemOzone $\downarrow$
    \\
 \hline\hline
 10\,$\text{kJ/m}^2$
    & 99.8\% & 98.7\% & 95\%
    & 68\%  \\
 \hline
 100\,$\text{kJ/m}^2$ 
    & 90\% & 60\% & 50\%
    & 91\%  \\ 
 \hline
 1000\,$\text{kJ/m}^2$ 
    & 25\% & 7\% & 4\%
    & 98\%  \\
 \hline\hline
\end{tabular}
\caption{The probability of at least one GRB in the given time, at the galactic radial position of the Earth, and the amount of the resulting ozone depletion. Data taken from \cite{Piran, Thomas}, combined for clarity.}
\label{table:grbs}
\end{table}


\section*{Who will Survive a GRB}
Our ultimate interest,
    in regard to the Fermi paradox,
    is in advanced complex life leading up (possibly) to intelligence and technology.
The evolutionary path is not linear along any clear path,
    and might not repeat in different circumstances.
We will thus follow the one known successful path,
    and try to estimate increased and reduced vulnerabilities along our own evolutionary history on Earth.
We will not be concerned with the death of any particular specimen or even the extinction on any particular species,
    but rather consider whether statistically,
    a given niche should, through Stages I \& II, preserve \emph{some} populations,
    to sustain \emph{some} species,
    of sufficient diversity and proximity to the relevant evolutionary path,
    that it may recover when Stage III is over, and carry on (not as before, but along a parallel path).

\subsection*{Stage I}
\begin{figure}[t]
\centering
\includegraphics[width=0.9\textwidth]{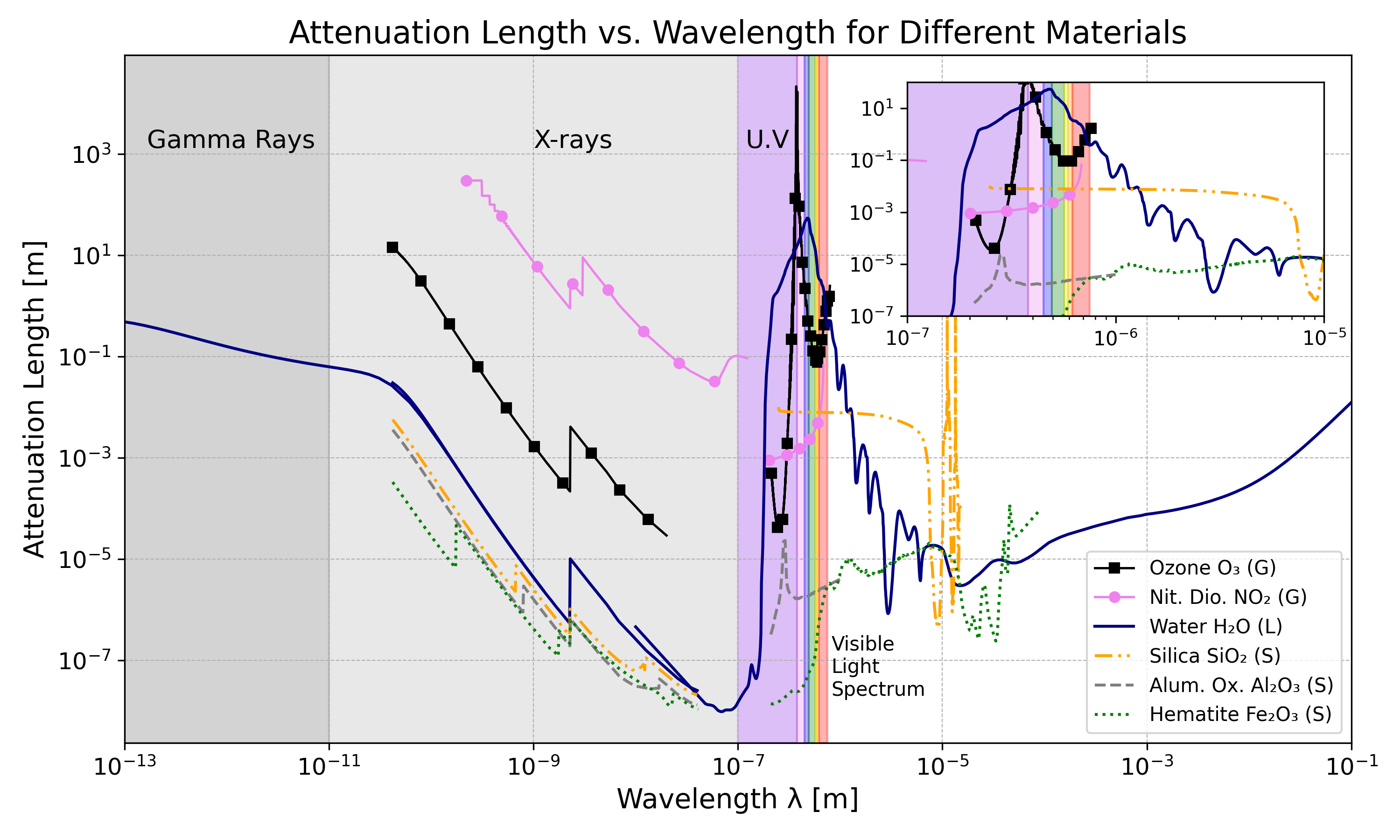}
\caption{\textbf{Attenuation of electromagnetic radiation through Earth-relevant environments.} X-ray data from CXRO \cite{Henke}; shortwave water data from NIST \cite{Hubbell}; UV/visible/IR for water from Segelstein \cite{Segelstein}; and for NO\textsubscript{2} from Schneider \cite{Schneider}. UV data for solid Al\textsubscript{2}O\textsubscript{3} and Fe\textsubscript{2}O\textsubscript{3} from Querry \cite{Querry}, SiO\textsubscript{2} from Herbin \cite{Herbin}. 
}
\label{fig:penetration}
\end{figure}

We first look more closely at the direct damage caused during $\gamma$-irradiation.
Paper II calculated that to radiatively damage extremophiles in the depths,
    the energy should suffice to boil the ocean.
However as is shown in Fig. \ref{fig:penetration},
    even in very shallow water (from mm to m), the gamma-ray intensity all but vanishes,
    while most visible light available for photosynthesis gets through.
Thus the focus on extremophiles is unnecessary -- complex life forms \emph{in general} are expected to survive Stage I underwater.
Underground gamma-ray shielding is even better,
    and photosynthesis is irrelevant.
We thus conclude that direct Stage I radiation is cataclysmically dangerous almost exclusively to life out of water and above ground.

To estimate the exposed land, we recall that GRB durations are very short relative to the Earth's rotation period.
While these durations can vary greatly,
    as do their fluxes\cite{Barnacka},
    still short GRBs (SGRBs) are all
    but over in under 2 seconds\cite{Mukherjee},
    while the average long GRB (LGRB) lasts about one minute \cite{Tarnopolski}.
These are all orders of magnitude shorter than the rotation period of the Earth, Mars, or even Jupiter,
    thus we can assume that the portion of the planet exposed to the GRB is only slightly over 50\% \cite{Dar}.
Even the 4-hour long GRB111209A \cite{Strattaa} would have covered only 67\% (80\%) of Earth (Jupiter) \cite{Russell},
    and such events are rare enough to be negligible statistically,
    so we maintain the 50\% (single hemisphere) as a representatve estimate.

Our planet's landmass is not equally distributed,
    and the distribution of land has not been constant over geological timescales.
We trace the probability of a hemispheric irradiation to have covered all (or even 90\%) of Earth's landmass since the beginning of the Phanerozoic era (from 541 Mya to the present) using Paleogeographic reconstructions \cite{Scotese,ScoteseLong} of Earth's continents.

We used the PaleoDEMS dataset \cite{Scotese}, which provides gridded reconstructions of Earth's continental surface at $1^{\circ} x  1^{\circ}$ resolution across the Phanerozoic eon (0-540 Mya).
For each time slice,
    land points were extracted
    using a land-sea mask.
Each land cell was converted into a 3D Cartesian vector on the unit sphere to facilitate geometric analysis.

\begin{figure}[t]
    \centering
    \begin{minipage}{0.32\textwidth}
        \centering
        \includegraphics[width=0.8\linewidth]{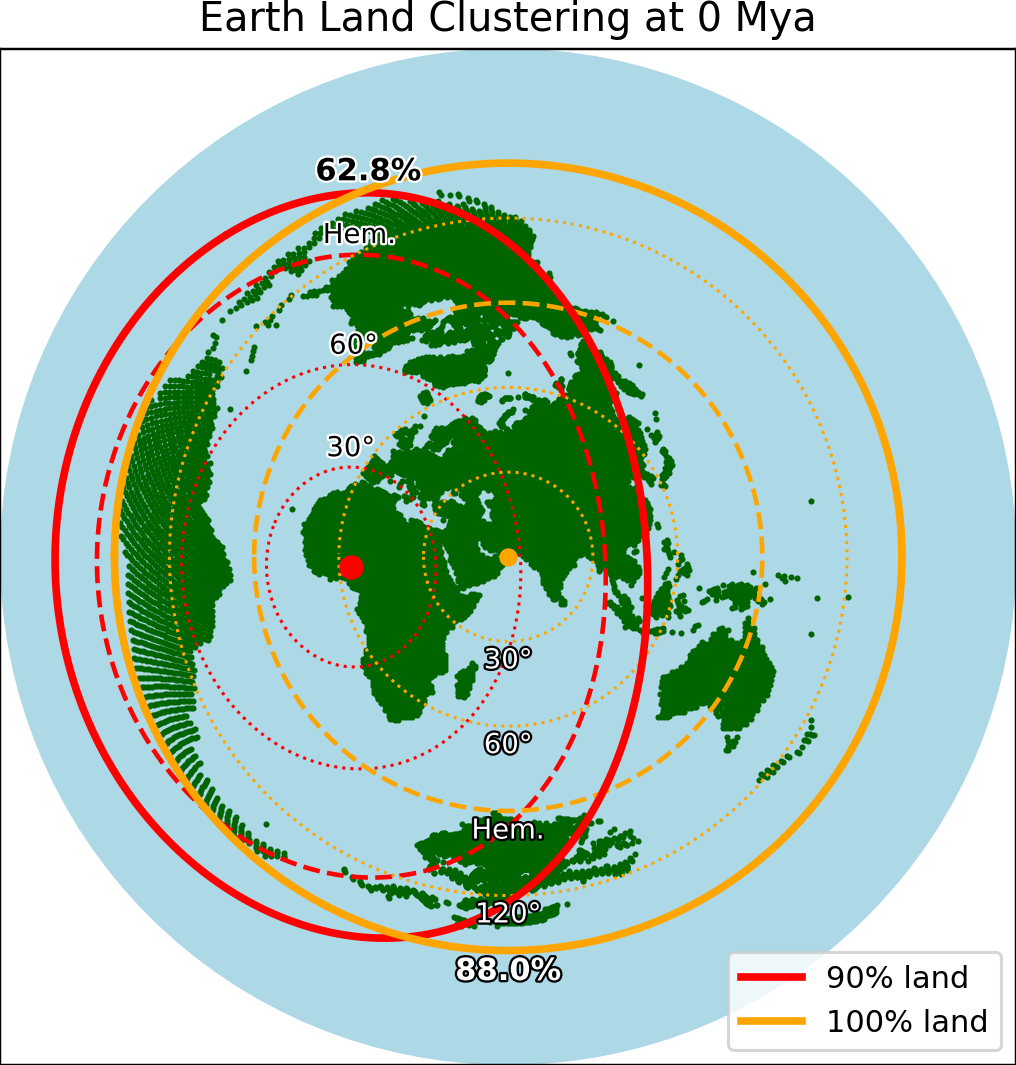}
    \end{minipage}\hfill
    \begin{minipage}{0.32\textwidth}
        \centering
        \includegraphics[width=0.8\linewidth]{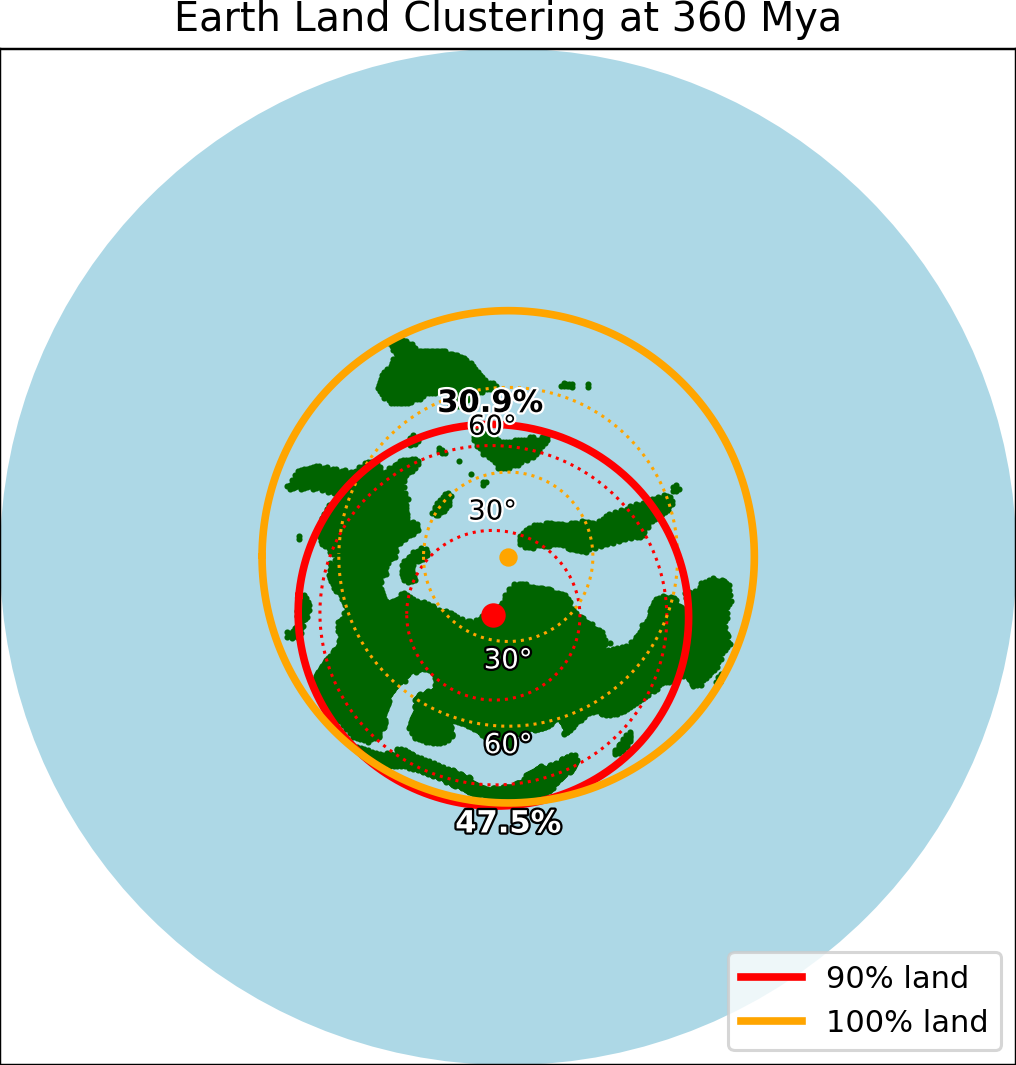}
    \end{minipage}\hfill
    \begin{minipage}{0.32\textwidth}
        \centering
        \includegraphics[width=0.8\linewidth]{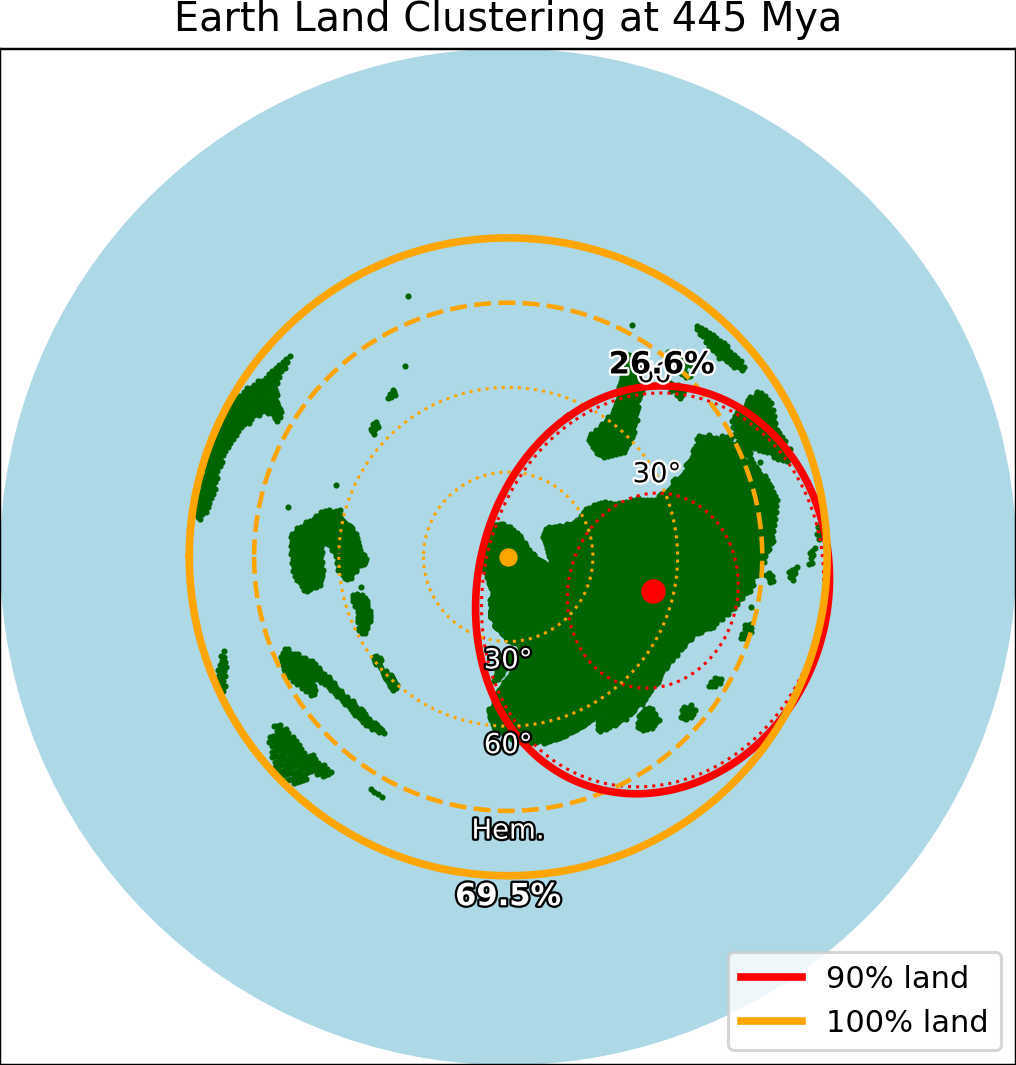}
    \end{minipage}
    \vspace{-5pt}
    \caption{Comparison of Earth's Azimuthal Equidistant projection (AEQD) maps at 0 Mya, 360 Mya, and 445 Mya.
    These refer to the clustering of Earth's land today,
        at maximal clustering for 100\% land,
        and at maximal clustering for 90\% land, respectively.
    Inner reference circles (small circles) are drawn at  fixed angular distances from the cap center for comparison. Figures written in black refer to the 90\% land data, figures written in white refer to the 100\% land data.}
    \label{fig:aeqd_maps}
\end{figure}

To evaluate GRB survivability,
    we developed \texttt{Python} code to compute the minimal spherical cap required to cover a given fraction of Earth's land at each time slice.
This cap represents the worst-case hemisphere exposed to a GRB.
The complementary hemisphere is assumed to be shielded from harmful Stage I flux,
    and the fraction of land it contains is used as a proxy for survivability.
We adopted a brute-force grid scan method,
    in which a dense set of test centers covering the entire globe (at $1^{\circ}$ intervals) was used to evaluate all possible spherical cap orientations.
For each test center,
    we calculated angular distances to all land points and identified the smallest angular radius $\theta$ such that a spherical cap centered at that point contained the desired land fraction (e.g., 90\% or 100\%).
The algorithm was implemented using \texttt{NumPy} and \texttt{Scikit-learn}'s \texttt{BallTree} with the haversine metric to enable fast nearest-neighbor queries on the sphere.
For each time step, we recorded:
\begin{itemize}
  \item The center and radius of the minimal cap enclosing the target land fraction.
  \item The fraction of land lying outside this cap (i.e., in the "safe" hemisphere).
  \item The dot--product between the cap center vector and the average position of the shielded land points, to evaluate alignment.
  Visual representation was also produced,
  examples of which appear in Fig. \ref{fig:aeqd_maps}.
\end{itemize}

Table \ref{table:land_coverage} summarizes the landmass clustering and dispersion,
    including the minimal angle $\theta$ and surface area fraction $f$ required for a round cap to cover 90\% (100\%) of the landmass (see Fig. \ref{fig:aeqd_maps}),
    and the probability $p$ of a single random GRB to irradiate that surface area.
For times when $\theta\geq\pi/2$, $p=0$;
    during denser clustering ($\theta\leq\pi/2$), $p = \left[1 - \sin(\theta)\right]/2$.
We find these to have been negligible,
    with $p_{90\%}\leq 5.8\%$ ($p_{100\%}\leq 0.06\%$),
    for the entire era,
    as can be seen in Fig. \ref{fig:oxygen_and_land}
We note this probability is governed by Earth's continental configuration \cite{LePichon},
    emphasizing a possible importance of plate tectonics.

We conclude that Stage I GRB damage is extremely unlikely to eradicate land-based life on a planet with similar land fractions and plate motions.

\begin{table}[H]
\centering
\caption{\textbf{Earth Land Clustering \& associated Risk across geological time.}
The table shows the relative surface area $f$ of the smallest spherical cap required to cover 90\% (100\%) of Earth's land.
The right column gives the chance of a GRB at a random sky angle to irradiate 90\% (100\%) of the land mass.
The data is derived from Paleogeographic reconstructions of the Earth over the last 540 Myr \cite{Scotese,ScoteseLong}.
We note that while the mean surface fraction since the Permian or Jurassic exceeds 50\%, the corresponding average chance of GRB coverage does not vanish, because the surface fraction and coverage probabilities are calculated per time step, only then averaged (and there were times when the surface fraction did dip slightly below 50\%).
}

\label{table:land_coverage}
\begin{tabular}{|c|c|c|c|}
\hline
\textbf{Land clustering/dispersion} &
\textbf{Geological time} & \textbf{Surface Fraction $f_{90\%} (f_{100\%})$} &
\textbf{GRB coverage $p_{90\%} (p_{100\%})$}
\\
\hline
Max. Clustering & 445 Mya (360 Mya)
    & 26.6\% (47.5\%)
    & 5.8\% (0.06\%)
\\
\hline
Avg.: since Phanerozoic & < 540 Mya - present >
    & 42.3\% (66.4\%)
    & 0.6\% (0.0001\%)
\\
\hline
Avg.: since Permian & < 300 Mya - present >
    & 50.1\% (69.6\%)
    & 0\% (0\%)
\\
\hline
Avg.: since Jurassic & < 200 Mya - present >
    & 54.8\% (76.2\%)
    & 0.05\% (0\%)
\\
\hline
Current & Present
    & 62.8\% (88\%)
    & 0\% (0\%)
\\
\hline
Max. Dispersion & Present (10 Mya)
    & 62.9\% (93.5\%)
    & 0\% (0\%)
\\
\hline
\end{tabular}
\end{table}

\subsection*{Stage II}
\begin{figure}[t]
\centering
\includegraphics[width=\textwidth]{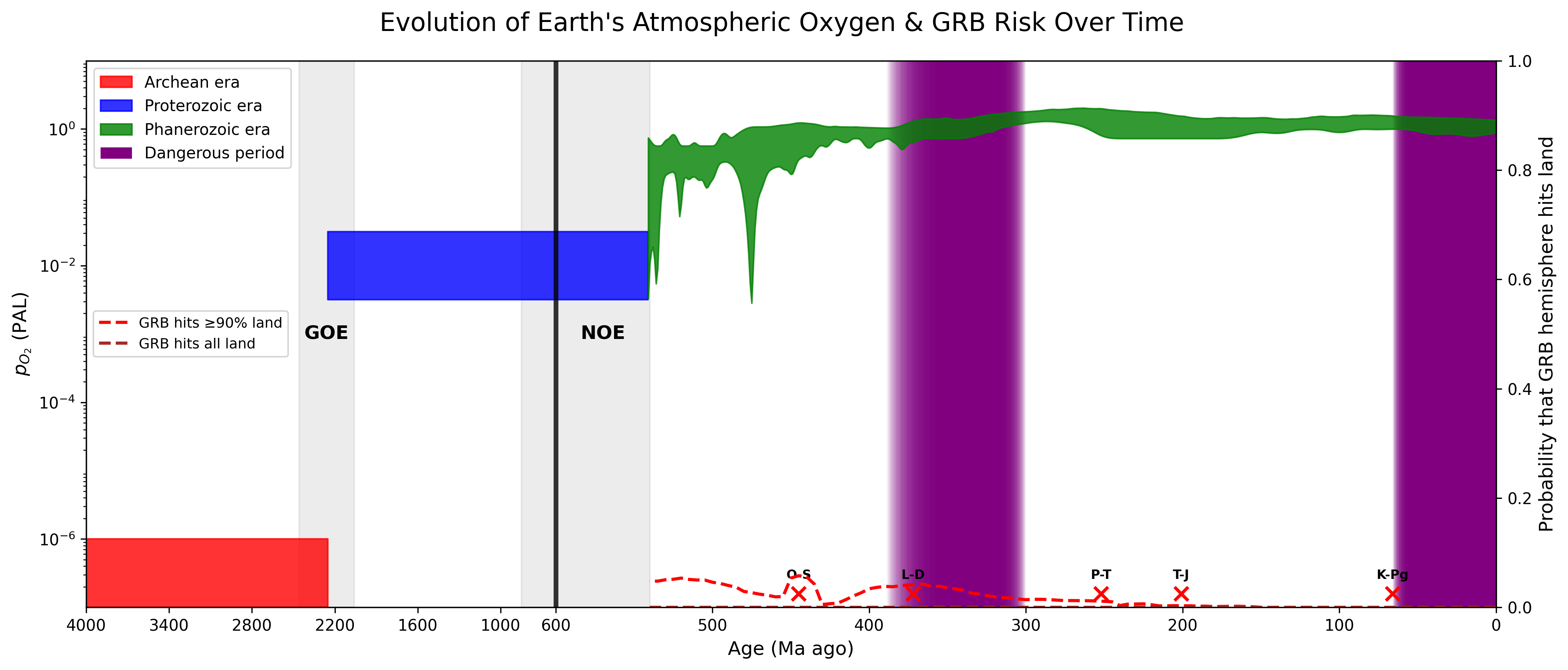}
\caption{\textbf{A timeline of Earth's evolution showing parameters relevant to GRB risks.} 
The horizontal axis is a broken linear axis: the left third spans 4000-–600 Mya (compressed), and the right two-thirds span 600-–0 Mya (expanded).
The oxygen data (thick, in three colors: red for Archaean, blue for Proterozoic, green for Phanerozoic) is from \cite{Lyons,Mills}, and follows the left-hand logarithmic concentration scale.
The thickness represents the uncertainty of the reconstructions.
The general oxygen trend has been that of increase, with shaded regions indicating major oxygenation events, i.e. the Great Oxidation Event (GOE \cite{Gumsley}) and the NOE \cite{Glass}.
The dashed red line (following the right-hand side axis) shows the probability of a potential GRB hemisphere encompassing at least 90\% of land, derived from paleogeographic reconstructions \cite{Scotese}.
This chance has had a generally downwards trend, and has been nil since the Jurassic/Cretaceous boundary.
Red X's mark the ``Big Five'' \cite{RaupSepkoski,Bond} extinction events.
We note that paleogeographic data exists further back \cite{PISAREVSKY2014207,ScoteseLong,MERDITH2021103477,Muller2022,LI2023104336,CAO2024101922},
	but in this work we are not concerned with such early land life.
The purple columns represent periods in our evolutionary lineage during which our ancestry was diurnal and hence at risk of UV exposure following a GRB;
    column edges fade to reflect uncertainty in the timing of these lifestyle transitions
    (see text for further details).
}
\label{fig:oxygen_and_land}
\end{figure}

It is thus the sustained exposure to UV radiation,
    rather than direct $\gamma$-irradiation, that poses the most danger to advanced life out of the water and above ground.
The first caveat we raise here is that the decrease in atmospheric Ozone can only crush life while life itself has become accustomed to its presence and shielding.
Prior to the accumulation of atmospheric oxygen, ozone depletion had not been a relevant stressor.
The Neoproterozoic Oxygenation Event (NOE\cite{Glass}),
    completed $\sim$550 Mya,
    established a persistent ozone layer,
    after which photochemical shielding against solar and cosmic UV became critical for surface habitability.
Before this interval,
    the absence of an ozone column rendered surface shielding largely irrelevant,
    with biota being predominantly aquatic or subterranean, inherently shielded from UV,
    or otherwise accustomed to it.
Oxygen (and Ozone) have thus only been relevant
    (at levels similar to today's, see Fig. \ref{fig:oxygen_and_land})
    during the Phanerozoic era,
    rendering the threat following a GRB (Stage II) at prior eras moot.
GRBs are thus unlikely to quench evolution  in the anaerobic eras.

Next we consider the aerobic era (the last $\sim\!\!540$ Myr).

Photochemical modeling indicates that a fluence of 100$kJ/m^2$ can deplete $>$90\% of the ozone column (see Table \ref{table:grbs}),
    maintaining elevated UVB and UVC irradiance at the surface for up to a decade \cite{Thomas}.
Under such conditions, persistence of significantly exposed taxa is improbable.
By contrast, a fluence of 10$kJ/m^2$ results in $\sim$68\% depletion,
    which is considered biologically irrelevant due to the presence of localized ``holes'' that facilitate recovery \cite{Thomas}.
Consequently,
    we adopt 100$kJ/m^2$ as the lower threshold for a GRB of evolutionary significance.
We note in Table \ref{table:grbs}
    (following Paper I)
    an estimate of a 50\% of such a GRB hitting the Earth over a 500 Myr period -
    effectively, over the entire Phanerozoic.

However, even for such intense GRBs and such depleted Ozone quantities,
    there is still effective shielding from UVB and UVC radiation provided by water\cite{Laurin} and by earth\cite{Dar,Frumkin} (Fig. \ref{fig:penetration}).
Our own evolutionary track should thus have been protected underwater until
    the emergence of tetrapods onto land $\sim\!\!350$ Mya \cite{Clack2009, Kumar2022},
    first raising exposure and vulnerability to increased UV during Stage II.
This is supported by phylogenomic and paleontological evidence placing our deep ancestry in entirely marine environments \cite{Telford2015, Swalla2008, Putnam2008, Robertson1957, Sallan2018, Zhu2022}.
We note that even previous conservative estimates,
    pushing the land transition to the time of 
    our shallow aquatic vertebrate ancestor \emph{Tiktaalik}\cite{Daeschler},
    only extend this period back to $\sim\!\!375$ Mya.
Thus the time-window of vulnerability is reduced by about 25\% relative to the shortest Table \ref{table:grbs} period.

Even later, our non-marine ancestors had mostly avoided spending considerable time exposed to the daylight sun,
    either through burrowing,
    restricting themselves to nocturnal activity,
    or both.
The reasons for these behaviours
    were not directly related to GRBs,
    but rather to the need of species which were sub-dominant and down the food chain
    to avoid being illuminated in full view of potential predators.
These evolutionary strategies
    and the adaptations associated with it,
    dubbed "the nocturnal bottleneck"\cite{Walls1942, Gerkema2013, Maor2017},
    have been associated with our ancestors for the period generally overlapping with the Mesozoic era (252–-66 Mya),
    i.e. the period during which dynosaurs dominated.
Divergence time analysis \cite{Kumar2022} combined with fossil evidence places the onset of synapsid nocturnality
    50 Myr earlier, at $\sim\!\!300$ Mya \cite{Angielczyk2014},
    with the nocturnal bottleneck persisting through the Mesozoic for all mammalian ancestors \cite{Maor2017}.

On the other end, the re-emergence of diurnal mammals has been associated with the time of the K-Pg mass extinction event\cite{RaupSepkoski,Bond},
    with estimates varying from about 75 Mya to 50 Mya,
    and with strict diurnality established in simian primates between 52--33 Mya \cite{Maor2017}.
Crucially, the same evolutionary strategies developed (or rather, niches exploited) by our ancestors to hide from predators by avoiding solar exposure,
    would have had the side-effect of also avoiding UV radiation after a GRB.
The vulnerability window thus shrinks to at most $\sim\!\!200$ Myr (375--250 Mya and 66 Mya--present), and as little as $\sim\!\!100$ Myr (350--300 Mya and 52 Mya--present).
We thus adopt the conservative value of $\sim\!\!200$ Myr as the relevant dangerous duration,
    since the NOE (i.e. less than 40\% of the time),
    and mark these windows in Fig \ref{fig:oxygen_and_land}.
This drastically decreases the chance of a GRB
    actually hitting inside the window(s).
We also note that these windows are at most $\sim\!\!100$ Myr and $\sim\!\!65$ Myr long (respectively),
    and thus estimate that if a GRB were to hit inside one of these windows,
    annihilating all diurnal species,
    the recovery (``fall-back'') time would still be around 100 Myr.

\section*{Mass Extinction Events}
A Phanerozoic-era GRB's ecological impact is still quite severe.
Following the NOE,
    organisms increasingly relied on atmospheric ozone for survival at the surface.
Not all surface organisms are nocturnal
    (following the previous reasoning, the \emph{dominant} species would \emph{not} be),
    and so mass extinction would surely follow such a biologically catastrophic perturbation \cite{Thomas}.
In addition, atmospheric \chemNOy production may induce opacity sufficient to reduce visible irradiance by up to 10\% \cite{Thomas},
    far exceeding the $\sim$0.36\% decline associated with the “Little Ice Age” \cite{Hoyt}.
Such reductions would amplify stress on photosynthetic primary producers,
    compounding the ecological disruption of Stage II (such an analysis is beyond the scope of the research presented here).
Such an extinction event would rank with "The Big Five"\cite{RaupSepkoski,Bond} major extinction events documented through Earth's Phanerozoic, namely (see Fig. 
  the Late Ordovician mass extinction (O-S) 445 Mya,
  the Late Devonian extinction/Kellwasser event (L-D) 372 Mya,
    the Permian–-Triassic extinction event (P-T) 252 Mya,
  the Triassic–-Jurassic extinction event (T-J) 201 Mya,
  and the Cretaceous–-Paleogene extinction event (K-Pg) 66 Mya.
The "Big Five" are notable for their extraordinary severity and their role in reshaping the trajectory of life on Earth. Each of these events corresponds to a sharp,
    identifiable drop in biodiversity,
    often recorded in the fossil record by abrupt changes in species diversity and abundance.
They mark boundaries between geological periods and are associated with major shifts in dominant life forms.
Likewise, a GRB would end many evolutionary lines - while also likewise, opening the way for others to have their opportunities.
Notably, a GRB during the Mesozoic era, i.e. the longest stretch of time during which our ancestors were nocturnal,
    could have wiped out their natural predators,
    and in fact hasten their diversification and take-over.

\section*{Conclusions: Re-Evaluating the Risk}

Paper I considered GRBs with a fluence above 100\,$\text{kJ/m}^2$ (during Stage I) to induce mass extinctions in Stage II -- and we accept this threshold.
Paper I then expected a region with an average rate of such GRBs above once per 500 Myr to be a no-go zone for life to develop and advance,
    due to successive quick eradications.
Paper I's mappings of these risk zones to our galaxy, using the stellar density,
    appear in the bottom half of Fig. \ref{fig:zones}.
For concreteness, Paper I also estimates that,
    at Earth's position in the galaxy,
    the probability of experiencing at least one GRB with a surface fluence of 100 kJ/m$^2$ in the last 500 Myr is approximately 50\%.

We have however shown that prior to about 540 Mya,
    Stage II damage would have been irrelevant (pre-NOE),
    and that for most of the time since then,
    our own ancestors were among the species least susceptible to daylight UV risk.
To re-evaluate the risk of a GRB,
    we ask what is the chance that a GRB (with a fluence above 100\,$\text{kJ/m}^2$) would hit during our ancestral exposure window(s),
    taken together to last at most 200 Myr, and as little as 100 Myr.
    While the 200 Myr window is not contiguous, for the strict purpose of Poissonian statistics that does not matter.

We use Paper I's Monte Carlo results,
    specifically
    their Fig. 4, providing
    the radially-resolved probability $P(\bar N > 1)$
    of having \emph{on average} more than one lethal GRB over a 1 Gyr period.
We approximate the lethal GRB rate at any radius as a Poissonian process,
    thus extract the rate $\lambda(r)$ from
    $P(r) = 1 - (1 + \lambda)\,e^{-\lambda}$
    (where over a period $t$, 
    $\bar N (r,t) = \lambda(r)\cdot(t/\text{Gyr})$).
For example,
    at Earth's galactic location this gives $\bar N\approx 0.88$ over 500 Myr,
    consistent with Paper I's quoted 50\% probability,
    while $\bar N \approx 0.35$ over 200 Myr.

We then map the galactic risk zones by the expected number of lethal GRBs $\bar N$
    over the relevant exposure window,
    defining three zones:
\begin{itemize}
    \item \textbf{High-risk}: $\bar N > 1.5$ -- more than one expected lethal GRB per window on average.
    \item \textbf{Moderate-risk}: $0.5 < \bar N < 1.5$ -- roughly one expected lethal GRB per window.
    \item \textbf{Low-risk}: $\bar N < 0.5$ -- less than one expected lethal GRB per window.
\end{itemize}

These risk zones are visualized in the upper half of Fig.~\ref{fig:zones},
    separately for the conservative (200 Myr) and aggressive (100 Myr) exposure window estimates,
    and are directly compared to Paper I's risk zones (Fig.~\ref{fig:zones}'s bottom half).
For 200 Myr, the high-risk boundary lies at $r \approx 1.2$ kpc and the moderate-risk boundary at $r \approx 5.5$ kpc;
    for 100 Myr these shrink further to $r \lesssim 0.6$ kpc (extrapolated) and $r \approx 2.2$ kpc respectively.
We note that the risky zones are drastically narrowed relative to Paper I,
    and in particular,
    that the Earth ($r \approx 8.5$ kpc) falls well within the low-risk zone under both window estimates,
    and thus no longer stands out as exceptional or ``lucky''.

\section*{Discussion}

Paper I considered a GRB every 500 Myr sufficient to stop Life,
    or at least its advanced evoution,
    and mapped the ensuing galactic danger zones.
Paper II considered the effects of such a GRB (or even stronger/closer) on the most extremely resilient known form of life,
    and concluded that it --
    and along with it, Life --
    will never go out on a planet (once started) due a random GRB from another star (other than the planet's own),
    effectively shrinking the galactic danger zones to naught.
We focused not on all Life nor on the most resilient species, but rather on the evolutionary history of \textit{homo sapiens} ourselves,
    and have found that througout Earth's history,
    our acnestors' susceptibility to extinction by GRBs had been much lower than implictly assumed --
    for concreteness,
    Stage II damage would have only been devastating over at most 200 of the last 540 Myr.
Our risk zones are thus not nullified,
    but much narrower relative to Paper I's.

Two opposite questions naturally arise,
    about a GRB hitting Earth during outside a window of exposure, or during a window of exposure, namely:
\begin{itemize}
    \item For a GRB hitting outside of an exposure window, would have our ancestors have survived the effets of Stage II on the entire collapsing eco-system, even if not directly killed off by UV radiation?
    \item Conversely, if a GRB had hit during the exposure window, would that have been the end of the evolutionary experiment on Earth, or rather, would evolutionary processes have recovered and re-developed advanced intelligent life (and how long would the process have been delayed)?
\end{itemize}
We make no attempt at an exact answer to these hypotheticals,
    but rather posit that as every mass extinction event has shown,
    the closing of a door on any species,
    opens doors to others.
The entire set of Life on Earth,
    at any given point,
    has had enough diversity to adapt,
    and diversification after extinction events have been very very fast,
    on the order of 10's of Myr.
We thus estimate that had our evolutionary line been severed,
    whether due to direct irradiation or to ecological collapse,
    another line would have survived, continued, and evolved.
Evolution has also demonstrated migration back and forth between niches,
    such as out of and back under the water,
    or out of and back into daylight.
Even eradicating all species since the last ``safe'' niche,
    the timeline shows a through-back time of no more than about 100 Myr.

Learning from the history of the Earth,
    we can thus assess that advanced evolution,
    leading up to land-based intelligence,
    may be stalled if strong enough GRBs occur more frequently than every 100-200 Myr on average (the range between the aggressive and conservative estimates).
Beyond hypothetical alternative Earth histories,
    these considerations are of interest regarding the possibilities and searches for Life elsewhere in our galaxy.
For our examination to be relevant,
    we must first note the assumptions regarding characteristics for such Earth-like planets,
    in addition to the usual habitable zone requirements.
These include being a rocky planet,
    with considerable surface oceans,
    orbiting a G-type star of Population I (for a similar radiation profile),
    with a day not too short and a year not too long,
    with a similar lifetime,
    and as noted previously,
    with similar tectonic behaviour,
    or otherwise similar land/water distribution.
In as much as such planets exist,
    we expect that throughout most of the galaxy (and in particular our own neighborhood),
    GRBs should not be a limiting factor.
In regions such as our own, then,
    the searches for signs of intelligence and technology need not be abandoned.
An open question for future research is whether biospheres which are in GRB-limited regions,
    dominated by subsurface or aquatic life,
    could be spectroscopically distinguishable from planets with thriving terrestrial biomes.


\section*{Data availability}
All paleogeographic reconstructions and absorption datasets used in this study are publicly available from the cited sources. Processed data and code can be shared upon request.

\section*{Research funding}
M.S. has been supported by the Israel Science Foundation (ISF) grant No. 1698/22, and is also thankful for the hospitality of the International Astronomical Union's Symposium (IAUS) 382 at Namour, Belgium, supported by an IAU grant.

\section*{Acknowledgements}
We also thank the Israel Physics Society (IPS) and the Israel Society for Astrobiology and the Study of the Origin of Life for opportunities to present and develop this work.
We are grateful to Dr. Gal Eyal (Bar-Ilan University) for his advice regarding the biological aspects of this research, and to Matan Mussel (Haifa University) and Gidi Yoffe (Weizmann Institute of Science) for helpful discussions.
Fig.~\ref{fig:stages} was designed by Lynn Maister (Shenkar College).
O.B. would also like to dedicate this work to the memory of fellow astrophysicist Noga (Hirshfeld) Sella, whose light shone very bright, and was extinguished much too soon, and to her family of survivors.

\section*{Author contributions statement}
M.~S. and O.~B. contributed equally, performing together all the research - but for the selection of species along evolutionary species, which A.~T. guided.

\section*{Competing interests}
The authors declare no competing interests.


\end{document}